\documentclass[lettersize,journal]{IEEEtran}
\IEEEoverridecommandlockouts

\usepackage{cite}
\usepackage[tbtags]{amsmath}
\usepackage{amssymb,amsfonts}
\usepackage{algorithm}
\usepackage{algorithmic}
\usepackage{graphicx}
\usepackage{textcomp}
\usepackage{xcolor}
\usepackage{caption}
\usepackage{commath}
\usepackage{multirow}
\usepackage{subcaption}
\usepackage{url}
\usepackage{gensymb}

\DeclareMathOperator*{\argmax}{arg\,max}
\DeclareMathOperator*{\argmin}{arg\,min}

\begin{document}

\title{Soft Handover Procedures in mmWave Cell-Free Massive MIMO Networks}

\author{Mahmoud Zaher, Emil Björnson, \IEEEmembership{Fellow, IEEE,} and Marina Petrova,\\ \IEEEmembership{Member, IEEE.}%
\thanks{Mahmoud Zaher, and Emil Björnson are with the Division of Communication Systems, KTH Royal Institute of Technology, 164 40 Stockholm, Sweden (e-mail: mahmoudz@kth.se; emilbjo@kth.se).}

\thanks{Marina Petrova is with the Division of Communication Systems, KTH
Royal Institute of Technology, 164 40 Stockholm, Sweden, and also with the Mobile Communications and Computing Group, RWTH Aachen University, 52062 Aachen, Germany (e-mail: petrovam@kth.se).}
\thanks{This work was supported by the FFL18-0277 grant from the Swedish Foundation for Strategic Research.}}

\maketitle

\begin{abstract}

This paper considers a mmWave cell-free massive MIMO (multiple-input multiple-output) network composed of a large number of geographically distributed access points (APs) simultaneously serving multiple user equipments (UEs) via coherent joint transmission. We address UE mobility in the downlink (DL) with imperfect channel state information (CSI) and pilot training. Aiming at extending traditional handover concepts to the challenging AP-UE association strategies of cell-free networks, distributed algorithms for joint pilot assignment and cluster formation are proposed in a dynamic environment considering UE mobility. The algorithms provide a systematic procedure for initial access and update of the serving APs and assigned pilot sequence to each UE. The principal goal is to limit the necessary number of AP and pilot changes, while limiting computational complexity. We evaluate the performance, in terms of spectral efficiency (SE), with maximum ratio and regularized zero-forcing precoding. Results show that our proposed distributed algorithms effectively identify the essential AP-UE association refinements with orders-of-magnitude lower computational time compared to the state-of-the-art. It also provides a significantly lower average number of pilot changes compared to an ultra-dense network (UDN). Moreover, we develop an improved pilot assignment procedure that facilitates massive access to the network in highly loaded scenarios.
\end{abstract}


\section{Introduction}

The wireless communication industry has seen tremendous growth over the past few decades. This ever-growing demand for wireless communication services poses a challenge for existing cellular networks. To satisfy the traffic demand, three key concepts have been widely investigated to increase the capacity of mobile networks: achieving higher spectral efficiency (SE), allocating more bandwidth, and deploying small cells (network densification) \cite{mendoza2020cluster}. Massive MIMO (multiple-input multiple-output) has been recognized as a cornerstone technology to enhance the performance of wireless communication systems by achieving higher SEs through spatial multiplexing of a large number of user equipments (UEs) on the same time-frequency resources \cite{marzetta2016fundamentals,bjornson2017book,nguyen2018optimal}. Despite the performance gain of massive MIMO, inter-cell interference and large pathloss variations remain inherent in the cell-centric design of conventional networks. Cell-free massive MIMO is a post-cellular architecture that synergistically combines cooperative principles with massive MIMO for seamless connectivity \cite{polegre2020channel}. A cell-free network is composed of a large number of distributed access points (APs) jointly serving the UEs within a given coverage area without creating cell boundaries \cite{bjornson2019making,chakraborty2020efficient,demir2021foundations,zaher2021learning}. It has become a basis for beyond 5G networks due to its ability to utilize macro diversity and higher resilience to interference \cite{guenach2020joint,demir2021foundations}, allowing for more network densification and providing almost uniform service to the UEs. Despite the gains provided by these technologies, the bandwidth limitation in sub-$6$ GHz bands remains a serious impediment in mobile networks. Combining cell-free massive MIMO with mmWave network operation can deliver unprecedented data rates with high reliability and improved network coverage.

A hindrance to the cell-free network topology is the computational complexity of signal processing and the fronthaul requirements for information exchange between the APs and central processing unit (CPU) \cite{zaher2021distributed}. To mitigate these issues and enable a scalable network operation, previous works have proposed user-centric (UC) clustering in which each UE is served by only a subset of the APs in the network \cite{bjornson2013book,buzzi2017cell,bjornson2020scalable}. This is deemed reasonable as the APs that are relatively far away from a given UE have a negligible impact on its performance \cite{mendoza2020cluster}. Pilot assignment and clustering of the APs to form the serving set for each UE can be based on a number of approaches. In \cite{buzzi2017cell}, AP clustering is such that each AP chooses to serve the UEs with the strongest estimated channels. Conversely, in \cite{mendoza2020cluster}, the clusters are formed such that each UE is served by $X$ APs that have the strongest channels assuming perfect channel knowledge, with the cluster size $X$ being varied for comparison. It is shown that even with small cluster size, comparable performance to the conventional form of cell-free networks can be achieved. The authors in \cite{polegre2020channel} propose two additional schemes for AP-UE associations alongside the two approaches mentioned earlier. The first scheme represents a variation of the dynamic cooperation clustering framework in \cite{bjornson2020scalable} where in addition to the APs choosing to serve the strongest UEs, each UE is appointed a \textit{Master} AP (that may or may not have chosen to serve that UE) to avoid situations where a UE is deprived of service by all APs. The second scheme is a combination of the two approaches in \cite{buzzi2017cell,mendoza2020cluster} to ensure that each UE will have a minimum number of serving APs as well as each AP will have a minimum number of UEs to serve. In situations where there are more UEs than orthogonal pilot sequences, the UEs were allowed to share the pilot sequences according to a pilot assignment procedure described in \cite{femenias2019cell}. An iterative joint power control and AP scheduling algorithm for the UL of a cell-free network was proposed in \cite{guenach2020joint}, where the algorithm alternates between optimizing the power coefficients assigned to the UEs by their serving APs and the AP-UE association matrix representing the serving set of APs to each UE until convergence is attained, while maintaining the assumption that the UEs are assigned orthogonal pilot sequences. Although such problem formulations may ultimately provide a better performance, the resulting high complexity potentially violates the real-time processing constraints that rely on resource scheduling and UE mobility \cite{zaher2021learning}.

As in the case of co-located massive MIMO, the performance of cell-free massive MIMO is critically affected by pilot contamination, especially when the number of UEs is significantly greater than the number of orthogonal pilot sequences and the UEs are non-uniformly distributed over the coverage area, resulting in a degraded quality of the acquired channel state information (CSI). The development of properly designed pilot assignment algorithms is thus pivotal to ensuring good performance in highly loaded networks \cite{buzzi2020pilot}. In \cite{buzzi2020pilot,d2020user}, the famous Hungarian algorithm \cite{kuhn1955hungarian,munkres1957algorithms} is utilized to find the optimal pilot assignment procedure either based on selected UC clusters or predetermined virtual clusters that a UE can be served by, respectively. Similarly, a multiple matching algorithm to find the optimal pilot assignment was developed in \cite{zhang2020noma}, with mobile users with similar covariance matrices being classified into the same cluster and allocated the same pilot. Non-orthogonal multiple access (NOMA) techniques were utilized to cancel the interference and reduce pilot contamination. In \cite{sabbagh2018pilot}, a dynamic pilot reuse scheme was developed such that a maximum of two UEs are allowed to share the same pilot sequence, under the condition that they have disjoint sets of serving APs. Additionally, pilot assignment schemes based on K-means clustering are proposed in \cite{attarifar2018random,chen2020structured}; where the former assumes all APs serve all UEs, while the latter considers that UEs are served by UC AP clusters. However, these schemes rely on maximizing the minimum geographical distance between UEs sharing the same pilots, thereby neglecting shadow fading effects. The effectiveness of such schemes is questionable in practical scenarios where obstructions in the propagation environment may lead to
tens of dBs of stronger/weaker average channel gains between the UEs and the APs that have the same geographical distance, which in turn affects the desired choice of pilot assignment. This is particularly evident for mmWave network operation, which is susceptible to severe shadowing upon line-of-sight (LOS) blockage. Moreover, these clustering schemes require some degree of centralized control \cite{attarifar2018random}, which increases the overhead in the network.

Moving towards mmWave cell-free networks, similar pilot assignment and cluster formation strategies are reported in \cite{alonzo2017cell,femenias2019cell}. An implication is that the objective relies on the AP cluster formation and pilot assignment strategy that ensure the right AP-UE associations and simultaneously reduce the effect of pilot contamination on the overall network performance. While all the mentioned algorithms are developed and tested in static conditions, the impact of changing the serving set of APs for each UE in a mobile network is not addressed. An initial study utilizing stochastic geometry to compare the performance of cell-free networks with that of UC clustering is presented in \cite{xiao2022mobility}. In addition, the authors in \cite{d2021user} present the first attempt for a handover algorithm in mmWave cell-free networks based on stochastic channel modeling without channel estimation, and accordingly only cluster formation and update were required with no developments to deal with pilot assignment/contamination.

In this paper, we address mobility management in the DL of a mmWave cell-free massive MIMO network. We consider fully digital beamforming based on imperfect CSI obtained from pilot training.\footnote{The analog beamforming in current mmWave systems adds an extra layer of complexity since analog beam sweeping must be done jointly with mobility management. However, we presume that digital mmWave transceivers will be the norm at the time when cell-free massive MIMO networks begin to be deployed.}
We notice that reliability and handover are particularly important in mmWave networks, where APs have short range.
Combining mmWave operation with cell-free massive MIMO essentially exploit spatial macro-diversity to reduce the occurrence of blocking events. This is fundamentally different from dual/multiple connectivity in 5G that allows connection to multiple APs on different frequencies. Aiming at extending traditional soft handover concepts to the more complicated AP-UE association strategies of cell-free networks, we develop distributed algorithms for joint pilot assignment and cluster formation in a dynamic environment considering UE mobility. The principal goal is to reduce the necessary number of AP and pilot sequence changes, and to develop distributed algorithms that facilitate massive access to the network in highly loaded scenarios. The proposed algorithms provide a systematic procedure for initial access as well as the update of the serving set of APs and assigned pilot sequence to each UE. An advantage of our proposed algorithms is that they do not require any kind of centralized control allowing for a fully distributed operation. To evaluate the performance of the proposed algorithms, we utilize ray-tracing software from \cite{simic2017demo} for mmWave network operation to obtain realistic and spatially consistent channel modeling environment. Further, we consider a typical deployment scenario for mmWave where low-power-low-tower (LPLT) APs are deployed along streets to serve mobile UEs. The main contributions of the paper can be summarized as follows:
\begin{itemize}
    \item We propose a distributed algorithm for joint pilot assignment and cluster formation that identifies the initial AP cluster and pilot assignment, where only UEs with disjoint AP clusters can share the same pilot, to reduce the level of pilot contamination. The algorithm is designed based on local decision metrics computed at a given AP or UE, to limit the signalling overhead by utilizing only the local information measured at each of them.
    
    \item We propose distributed algorithms for dynamic pilot assignment and cluster update, considering UE mobility, to extend traditional handover concepts to the more complicated AP-UE associations of the cell-free network topology. The algorithm is tailored to minimize \textit{Master} AP handovers and necessary pilot assignment decisions, and to facilitate massive access to the network. The algorithm relies on local decision metrics computed at a given AP or UE using only the local information measured at each of them. Further, we perform a detailed complexity analysis of the proposed algorithms and compare them to some relevant designs in the literature.
    
    \item We propose a modified pilot assignment strategy that takes into account both the average channel gains and the load distribution (i.e., used pilot sequences) of the possible serving APs that are in the vicinity of a given UE, showing improved performance, especially in highly loaded scenarios. The algorithm balances between the interference measured on a given pilot at the \textit{Master} AP and the availability of other APs near the UE.
    
    \item We develop a \textit{Maneuvering Smooth Random Waypoint Mobility} model to evaluate the performance of our proposed algorithms with realistic random UE mobility in a site map including buildings and other obstructions. We stress that existing waypoint mobility models are not suitable for simulating UE mobility where obstructions exist in the mobility path. This is particularly important for accurate channel modeling in mmWave scenarios that are susceptible to severe shadowing upon LOS blockage.
\end{itemize}

The rest of the paper is organized as follows: Section~\ref{system} outlines the system model. The joint pilot assignment and AP cluster formation algorithm for initial access is described in Section~\ref{pilot}. Section~\ref{handover_sec} details the proposed soft handover procedures. Section~\ref{SSB} presents the modified pilot assignment strategy. Section~\ref{numerical} presents numerical results, whereas the main conclusions of the paper are stated in Section~\ref{conc}.

\textbf{Notations:} Lowercase and uppercase boldface letters denote column vectors and matrices, respectively. The symbols $(\cdot)^*$, $(\cdot)^T$, and $(\cdot)^H$ indicate conjugate, transpose and conjugate transpose, respectively. $\mathbb{E}(\cdot)$, tr$(\cdot)$, and $\norm{\cdot}$ denote the expectation, trace, and L$_2$ vector norm, respectively. $\mathbf{I}_M$ represents the $M \times M$ identity matrix.

\section{System Model}\label{system}

In this paper, we consider a cell-free massive MIMO network with $L$ APs, each equipped with $N$ antennas, and $K$ single-antenna UEs arbitrarily distributed in a large service area. To achieve a scalable network operation, each UE is jointly served by only a subset of the APs in the network. The standard block fading channel model is assumed where the time-varying wideband channels are divided into time-frequency coherence blocks such that the channels are static and frequency-flat in each block \cite{bjornson2017book}. The coherence block consists of $\tau_c$ symbols. As the UEs move, the channel coefficients are updated following the update of the UE location and the geometry of the site map. We utilize a ray-tracing software \cite{simic2017demo}, suitable for mmWave frequency operation, that provides realistic and spatially consistent channel realizations. The ray-tracer employs specular reflections upon the different objects in the site map to determine the path gains from AP $l$, $l = 1, \dots, L$, that are received at a given location. The path gains are combined to create the channel coefficients in each coherence block. The channel between UE $k$ and AP $l$ is represented by $\mathbf{h}_{kl} \in \mathbb{C}^{N \times 1}$ and is given by
\begin{equation}
\mathbf{h}_{kl} = \sum_{i = 1}^{N_{kl}}\alpha_i\mathrm{e}^{-j\psi_i}\mathbf{a}\hspace{-1pt}\left(\phi_i\right),
\end{equation}
where $\alpha_i$ corresponds to the gain of path $i$, $\psi_i \sim \mathcal{U}[0, 2\pi)$, $\mathbf{a}\hspace{-1pt}\left(\phi_i\right)$ is the array response vector of AP $l$ with $\phi_i$ being the angle of departure (AoD) of the $i^{th}$ path\footnote{Typically, the array response vector depends on both the azimuth and elevation angles, however, since we consider horizontally placed linear arrays, $\mathbf{a}\hspace{-1pt}\left(\cdot\right)$ is a function of the azimuth AoD $\phi_i$ only and is rotation invariant in the elevation plane.}, and $N_{kl}$ represents the number of paths from AP $l$ to the location of UE $k$. We consider a 3D environment where a uniform linear array (ULA) at each AP is horizontally placed with half-wavelength spacing between the antenna elements, such that
\begin{equation}
\mathbf{a}\hspace{-1pt}\left(\phi_i\right) = [1 ~ \mathrm{e}^{-j\pi\sin{\phi_i}} ~ \mathrm{e}^{-j2\pi\sin{\phi_i}} ~ \hdots ~ \mathrm{e}^{-j\left(N-1\right)\pi\sin{\phi_i}}]^T.
\end{equation}

\begin{figure}
\centering
\setlength{\abovecaptionskip}{0.3cm}
\includegraphics[scale=0.56]{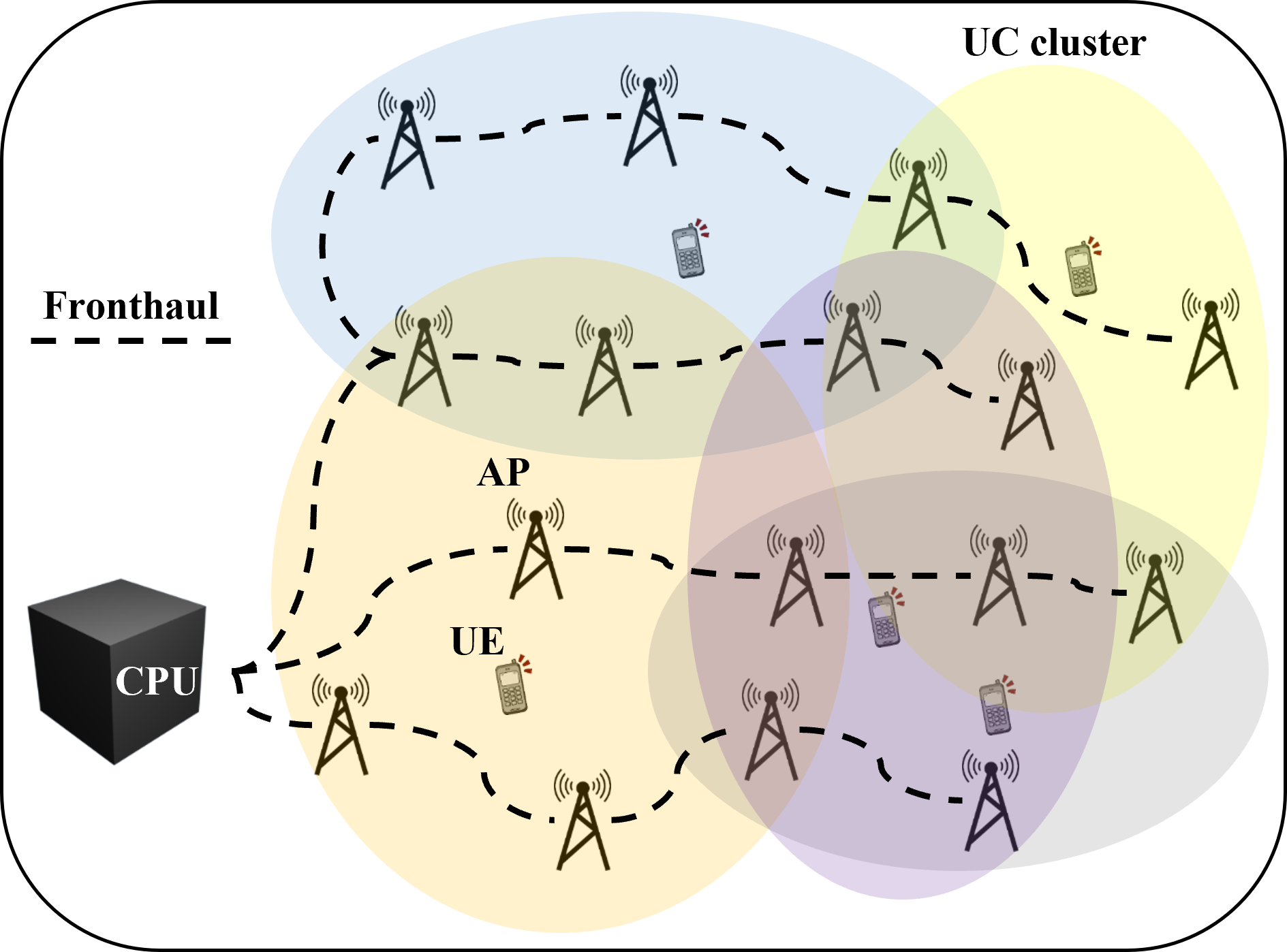}
\caption{User-centric cell-free massive MIMO. Each UE is served by a subset of the APs.}
\label{model1}
\vspace{-1em}
\end{figure}

The average channel gain from a given antenna at AP $l$ to UE $k$ is determined by $\beta_{kl} = \sum_{i = 1}^{N_{kl}}\alpha_i^2$. The network architecture is depicted in Fig.~\ref{model1}. Fronthaul links connect the APs to a CPU, conveying UL and DL data between them and other necessary signals. Primarily, such a network can be operated in two ways: centralized and distributed operation \cite{demir2021foundations}. The distributed operation is adopted in this paper, where each AP selects the precoding vectors based on the locally estimated channels. The connections are assumed to be error-free, and no instantaneous CSI is transmitted over the fronthaul, but only data and channel statistics  \cite{bjornson2019making,nayebi2017precoding}. 

\subsection{Channel Estimation}

A time-division duplex (TDD) protocol comprised of a pilot transmission phase for channel estimation and a data transmission phase is considered. As in \cite{demir2021foundations}, each coherence block is divided into three parts: $\tau_p$ symbols for UL pilots, $\tau_u$ symbols for UL data transmission, and $\tau_d$ symbols for DL data transmission. Accordingly, the coherence block is written as $\tau_c = \tau_p + \tau_u + \tau_d$. Since this paper focuses on mobility management in the DL of a cell-free network, only UL pilot and DL data transmission are considered, i.e., $\tau_u=0$.

In the channel estimation phase, the UEs are each assigned a $\tau_p$-length pilot from a set of $\tau_p$ mutually orthogonal pilot sequences utilized by the APs. We consider the general case that $\tau_p < K$, and propose a joint pilot assignment and AP cluster formation algorithm to handle initial access to the network as well as the update of the serving set of APs and assigned pilot to each UE considering UE mobility. The details of the algorithms are described in Sections~\ref{pilot} and~\ref{handover_sec}.

Let $\mathcal{P}_t \subset \{1, \hdots, K\}$ and $\boldsymbol{\theta}_{t_k} \in \mathbb{C}^{\tau_p \times 1}$ denote the set of UEs that are assigned to pilot $t$ and the  pilot sequence assigned to UE $k$, respectively. The pilot sequence $\boldsymbol{\theta}_{t_k}, \forall k$ is assigned from a set of $\tau_p$ mutually orthogonal pilot sequences available at the APs. The received signal at AP $l$, which is intended for the estimation of the UEs' channels, $\mathbf{Y}_l^p \in \mathbb{C}^{N \times \tau_p}$ is computed as
\begin{equation}
\mathbf{Y}_l^p = \sum_{i = 1}^K\sqrt{p_i}\mathbf{h}_{il}\boldsymbol{\theta}_{t_i}^T + \mathbf{N}_{l},
\end{equation}
where $p_i\hspace{-1pt}$ is the transmit power of UE $i$ and $\mathbf{N}_{l} \in \mathbb{C}^{N \times \tau_p}$ represents the additive i.i.d.~complex Gaussian noise at AP $l$ during the entire channel estimation phase. We cannot take the conventional Bayesian estimation approach \cite{demir2021foundations} since it requires stationary channel statistics, which are not obtained with our ray-traced channel realizations and substantial UE mobility. Instead, we will utilize a least square (LS) estimator. Hence, the channel between UE $k \in \mathcal{P}_{t_k}$ and AP $l$ is estimated as
\begin{equation}
\hat{\mathbf{h}}_{kl} = \frac{\mathbf{Y}_l^p\boldsymbol{\theta}_{t_k}^*}{\tau_p\sqrt{p_k}} = \mathbf{h}_{kl} + \sum_{i \in \mathcal{P}_{t_k}\backslash\{k\}}\hspace{-5pt}\sqrt{\frac{p_i}{p_k}}\mathbf{h}_{il} + \frac{\mathbf{n}_{t_kl}}{\sqrt{\tau_pp_k}},
\label{estimate}
\end{equation}
where the second term represents the pilot contamination due to pilot sharing UEs with UE $k$ and $\mathbf{n}_{t_kl} =\frac{1}{\sqrt{\tau_p}}\mathbf{N}_{l} \boldsymbol{\theta}_{t_k}^* \hspace{-0.85pt} \sim \mathcal{N}_{\mathbb{C}}\hspace{-0.5pt}\left(\mathbf{0}, \sigma^2\mathbf{I}_N\hspace{-0.5pt}\right)$ is the complex Gaussian noise vector at AP $l$ after correlating the signal with pilot $t_k$. Note that the channel estimates provided are only between a UE and its serving cluster of APs, i.e., each AP estimates only the channels of the UEs that are being served by that AP.

\subsection{Downlink Data Transmission}

In the data transmission phase, each UE is assumed to be served by only a subset of the $L$ APs in the network; for UE $k$ that is, $\mathcal{M}_k \subset \{1, \hdots, L\}$. As a result, the received DL signal at UE $k$ is given by
\begin{equation}
y_k^{\textrm{dl}} = \sum_{l = 1}^{L}\mathbf{h}_{kl}^H\sum_{i = 1}^{K}\sqrt{\rho_{il}}\mathbf{D}_{il}\mathbf{w}_{il}s_i + n_k
\end{equation}
where $\rho_{il}\geq 0$ is the allocated DL power by AP $l$ to UE $i$ and $\mathbf{w}_{il} \in \mathbb{C}^{N \times 1}$ represents the corresponding normalized precoding vector, such that $\norm{\mathbf{w}_{il}}^2 = 1$. In addition, $s_i$ denotes the zero-mean signal intended for UE $i$ and $n_k \sim \mathcal{N}_{\mathbb{C}}\hspace{-1pt}\left(0, \sigma^2\right)$ represents the noise at UE $k$. $\mathbf{D}_{il}$ determines whether AP $l$ serves UE $i$ such that the effective transmit precoding vector $\mathbf{D}_{il}\mathbf{w}_{il}$ is given by
\begin{equation}
\mathbf{D}_{il}\mathbf{w}_{il} =
\begin{cases}
\mathbf{w}_{il} & \textrm{for } l \in \mathcal{M}_i, \\
\mathbf{0} & \textrm{for } l \notin \mathcal{M}_i.
\end{cases}
\end{equation}

Employing the use-and-then-forget (UatF) capacity bounding method \cite{bjornson2017book} and similar to \cite{chakraborty2020efficient}, a lower bound on the ergodic DL capacity over fast-fading channels with imperfect CSI in a cell-free system can be obtained. It is called an achievable SE and the SE expression for UE $k$ is 
\begin{equation}
\textrm{SE}_k = \frac{\tau_d}{\tau_c}\textrm{log}_2\left(1 + \textrm{SINR}_k\right) \label{eq:SEk} \quad \textrm{bit/s/Hz}
\end{equation}
where
\begin{equation}
\textrm{SINR}_k = \frac{\left(\mathbf{a}_k^T\boldsymbol{\mu}_k\right)^2}{\sum_{i = 1}^{K}\boldsymbol{\mu}_i^T\mathbf{B}_{ki}\boldsymbol{\mu}_i - \left(\mathbf{a}_k^T\boldsymbol{\mu}_k\right)^2 + \sigma^2}
\label{SINR}
\end{equation}
represents the effective signal-to-interference-and-noise ratio (SINR) and
\begin{align}
&\boldsymbol{\mu}_k = \left[\mu_{k1} \cdots \mu_{kL}\right]^T \in \mathbb{R}^{L \times 1},\ \mu_{kl} = \sqrt{\rho_{kl}} \label{eq:mu}\\
&\mathbf{a}_k = \left[a_{k1} \cdots a_{kL}\right]^T \in \mathbb{R}^{L \times 1},\ a_{kl} = \mathbb{E}\left\{\mathbf{h}_{kl}^H\mathbf{D}_{kl}\mathbf{w}_{kl}\right\}\\
&\mathbf{B}_{ki} \in \mathbb{R}^{L \times L},\ b_{ki}^{lm} = \Re\left(\mathbb{E}\left\{\mathbf{h}_{kl}^H\mathbf{D}_{il}\mathbf{w}_{il}\mathbf{w}_{im}^H\mathbf{D}_{im}^H\mathbf{h}_{km}\right\}\right)
\end{align}
and $b_{ki}^{lm}$ corresponds to element ($l$, $m$) in matrix $\mathbf{B}_{ki}$. The SE value depends on what CSI is available, but as with any other SE expression, it can be achieved without error.
\vspace{2pt}

In \eqref{SINR}, the term $\mathbf{a}_k^T\boldsymbol{\mu}_k = \sum_{l=1}^{L} \sqrt{\rho_{kl}} \mathbb{E}\left\{\mathbf{h}_{kl}^H\mathbf{D}_{kl}\mathbf{w}_{kl}\right\}$ in the numerator represents the desired signal gain for UE $k$ over deterministic precoded channel generated by the UatF bound. The SE expression is applicable for any precoding scheme, with precoding vectors rotated such that $a_{kl} = \mathbb{E}\left\{\mathbf{h}_{kl}^H\mathbf{D}_{kl}\mathbf{w}_{kl}\right\}$ is real and non-negative. The precoding vectors $\{\mathbf{w}_{il}\}$ satisfy short-term power constraints, i.e., $\norm{\mathbf{w}_{il}}^2 = 1$ is satisfied in each coherence block. The normalized precoding vector is defined as $\mathbf{w}_{il} = \bar{\mathbf{w}}_{il}/\norm{\bar{\mathbf{w}}_{il}}$, such that $\bar{\mathbf{w}}_{il}$ can be selected arbitrarily.

In the numerical evaluation, we will employ the maximum ratio (MR) and local partial regularized zero-forcing (RZF) precoding schemes. The precoding vectors are thus defined as
\begin{equation}
\bar{\mathbf{w}}_{kl} =
\begin{cases}
      \hat{\mathbf{h}}_{kl} & \textrm{for MR}, \\
      \left(\sum\limits_{i \in \mathcal{D}_l}p_i\hat{\mathbf{h}}_{il}\hat{\mathbf{h}}_{il}^H + \sigma^2\mathbf{I}_N\right)^{-1}p_k\hat{\mathbf{h}}_{kl} & \textrm{for RZF},
\end{cases}
\end{equation}
where $\mathcal{D}_l \subset \{1, \hdots, K\}$ denotes the set of UEs served by AP $l$. Note that the main contributions of this work are applicable along with any other precoding scheme as well.

Moreover, we adopt a low-complexity heuristic power allocation inspired by \cite{zaher2021distributed}, however, considering only the set of UEs served by a given AP. Hence, the allocated  DL power coefficients for AP $l$ are given by
\begin{equation}
\rho_{kl} = P_{\textrm{max}}^{\textrm{dl}}\frac{\left(\beta_{kl}\right)^v}{\sum\limits_{i \in \mathcal{D}_l}\hspace{-2pt} \left(\beta_{il}\right)^v}, \quad k \in \mathcal{D}_l,
\label{input}
\end{equation}
where $P_{\textrm{max}}^{\textrm{dl}}$ is the maximum per-AP power budget and $v$ is a constant exponent that reshapes the large-scale fading coefficients. Note that \eqref{input} represents the state-of-the-art heuristic allocation and can be implemented in a distributed manner \cite[Sec.\,7.2.3]{demir2021foundations}. On the other hand, optimized power allocation requires centralized computations which stands in contrast to our adoption of distributed network operation where several functionalities are implemented at the AP level with information readily available at the AP. Such distributed operation reliefs the fronthaul requirements for performing these functionalities. We stress that the system model is applicable with any mobility model used to update the UE locations. In this work, we develop a Maneuvering Smooth Random Waypoint Mobility model, detailed in Section \ref{mobility}, that generates realistic random UE mobility on a map including obstructions for evaluation of the proposed algorithms.

\section{Initial Pilot Assignment and Cluster Formation}\label{pilot}

As a prerequisite to the proposed \textit{Pilot Assignment and Cluster Update} algorithm, we define another algorithm for \textit{Initial Access}. The initial access part determines the first pilot assignment and serving set of APs to all UEs, which then must be revised as the UEs move around in the area of interest. Hence, it serves as an initialization for the proposed pilot and cluster update algorithms.
Based on each UE's speed, we can determine how quickly the average channel gains may change to the extent that the preferred AP-UE associations must be updated. As a result, we need to make a decision once per such interval, denoted by $T_i$, whether to revise the serving APs and assigned pilot to each UE, i.e., $T_i$ represents the simulation sampling time. For brevity, we introduce a discrete time index $n$ to describe variables' update at each interval.

To detail the proposed algorithms, we first define $M_{\mathrm{max}}$ as the maximum AP cluster size serving any given UE, $M_{k,\mathrm{min}}^{\left(n\right)}$ and $N_l^{\left(n\right)}$ as the number of APs having noticeable average channel gains to UE $k$ and the number of UEs having noticeable average channel gains to AP $l$ that are above the noise floor at time instant $n$, respectively. Accordingly, $M_{k,\mathrm{min}}^{\left(n\right)}$ and $N_l^{\left(n\right)}$ are determined based on the received signal strengths at a given UE or AP, whereas $M_{\mathrm{max}}$ is a tunable parameter that can be set on the radio network controller (RNC) level. Moreover, we define the sorting operator $O^{\left(n\right)} : \left\{1, \hdots, x\right\} \rightarrow \left\{1, \hdots, x\right\}$ that sorts entries in descending order based on the average channel gains between APs and UEs.

The initial pilot assignment and cluster formation is devised in Algorithm~\ref{alg1}. In the first part, Steps~\ref{part1} through \ref{part1end}, a temporary set composed of UEs having the strongest average channel gains is formed for each AP $\mathcal{D}_l\left[0\right], l = 1, \hdots, L$, with a cardinality of at most $\tau_p$ UEs (so different pilots could be allocated). Afterwards, each UE attempts to connect to the AP with the strongest average channel gain, say AP $l$. If the UE does not belong to the set of UEs with the strongest signal formed for AP $l$, that is $\mathcal{D}_l\left[0\right]$, the request is rejected. The UE then keeps on attempting to connect to an AP in descending order of perceived average channel gain between the UE and all APs in its vicinity (Steps~\ref{part2} through \ref{part2end}), i.e., the APs with non-negligible average channel gains to the UE. If the UE fails to connect to all the APs in its vicinity, the UE is denied access to the network. Otherwise, the first AP to accept the connection request is appointed as the \textit{Master} AP for that UE (Step~\ref{master_appoint}). In Step~\ref{pilot_int}, the pilot sequence, among the set of free pilots at the master AP $\mathcal{T}_{l_k^{\star}}$, where the master AP measures the least interference is assigned to the UE as per \eqref{best_pilot}. Note that especially in lightly loaded scenarios, an AP may not sense any interference on several pilot sequences. In this case, a pilot sequence is chosen uniformly at random from the set of pilots with no perceived interference, such that the algorithm does not favour the assignment of a specific pilot sequence over others. Next, in Steps~\ref{report} and \ref{invitation}, UE $k$ reports the set of APs with the largest average channel gains $\mathcal{A}_k\left[0\right]$, starting from $O_{k}^{\left(0\right)}\left(i+1\right)$, to its master AP. The master AP then sends a service invitation to the reported APs by the UE.

A limitation to the performance of a cell-free network is the level of pilot contamination, because it reduces the channel estimation quality at the APs and creates coherent interference \cite{demir2021foundations}.
To limit the pilot contamination perceived in \eqref{estimate}, each AP is allowed to serve only one UE per pilot. As a result, the maximum number of UEs that can be simultaneously served by an AP is limited to $\tau_p$.
Consequently, in Step~\ref{1UEperpilot} an AP in $\mathcal{A}_k\left[0\right]$ accepts the service invitation if and only if it has the pilot assigned to UE $k$ free. Note that this does not completely eliminate the interference in the channel estimation phase. The reason is that two UEs having disjoint AP clusters may share the same pilot sequence, which does not necessarily imply that the channel gains between the UEs and each other's AP clusters are zero. We highlight that the procedure described in Algorithm~\ref{alg1} features a solution to the multiple matching problem in a static scenario, where the AP clusters are formed by matching the $\min(\tau_p,N_l^{(0)})$ strongest UEs for service by any given AP and the $\min(M_{\mathrm{max}},M_{k,\mathrm{min}}^{(0)})$ strongest APs to any given UE, conditioned on the availability of the pilot sequence, appointed by the master AP, at the other APs.

\begin{algorithm}[t]
    \caption{Initial Pilot Assignment and Cluster Formation}
    \label{alg1}
    \noindent\textbf{Input:} Initial average channel gains $\beta_{kl}\left[0\right] \geq 0$ between AP $l$ and UE $k, \forall l, k$.
    \begin{algorithmic}[1]
        \FOR{$l = 1, \hdots, L$}\label{part1}
        \STATE Sort at most the $\tau_p$ UEs with highest average channel gains $\beta_{kl}\left[0\right], k = 1, \hdots, K$.
        \STATE Set $\mathcal{D}_l\left[0\right] \gets  \left\{O_{l}^{\left(0\right)}\left(1\right), \hdots, O_{l}^{\left(0\right)}\left(\min\left(\tau_p,N_l^{\left(0\right)}\right)\right)\right\}$.
        \ENDFOR \label{part1end}
        \FOR{$k = 1, \hdots, K$}
        \STATE Sort the APs in descending order based on $\beta_{kl}\left[0\right], l = 1, \hdots, L$.
        \STATE Set $\mathcal{A}_k\left[0\right] \gets \left\{O_{k}^{\left(0\right)}\left(1\right), \hdots, O_{k}^{\left(0\right)}\left(M_{k,\mathrm{min}}^{\left(0\right)}\right)\right\}$.
        \STATE Set $i \gets 1$.
        \WHILE{$k \notin \mathcal{D}_{O_k^{\left(0\right)}\left(i\right)}\left[0\right] \And i \leq M_{k,\mathrm{min}}^{\left(0\right)}$}\label{part2}
        \STATE UE $k$ sends a connection request to AP $O_k^{\left(0\right)}\left(i\right)$.
        \STATE AP $O_k^{\left(0\right)}\left(i\right)$ rejects service to UE $k$.
        \STATE Set $i \gets i + 1$.
        \STATE Set $\mathcal{A}_k\left[0\right] \gets \mathcal{A}_k\left[0\right]\backslash\{O_{k}^{\left(0\right)}\left(i\right)\}$.
        \ENDWHILE \label{part2end}
        \IF{$i \leq M_{k,\mathrm{min}}^{\left(0\right)}$}
        \STATE UE $k$ sends a connection request to the next AP with the strongest signal, $O_k^{\left(0\right)}\left(i\right)$.
        \STATE $O_k^{\left(0\right)}\left(i\right)$ is assigned as the \textit{Master} AP of UE $k$ and denoted as $l_k^{\star}$. \label{master_appoint}
        \STATE \label{pilot_int} $l_k^{\star}$ assigns the pilot sequence which it senses the least interference on to UE $k$; that is \vspace{-2pt} \begin{equation}
            t_k = \argmin_{t \in \mathcal{T}_{l_k^{\star}}} \sum_{i \in \mathcal{P}_t}\beta_{il_k^{\star}}\left[0\right].
            \label{best_pilot}
        \end{equation}
        \STATE \label{report} UE $k$ reports the set $\mathcal{A}_k\left[0\right]$ to $l_k^{\star}$.
        \STATE \label{invitation} $l_k^{\star}$ sends a service invitation on pilot $t_k$ to the reported APs by UE $k$.
        \STATE \label{1UEperpilot} An AP in $\mathcal{A}_k\left[0\right]$ accepts invitation $\iff$ Pilot $t_k$ is free on that AP.
        \STATE \label{M_ksetting} Set $\mathcal{M}_k\left[0\right] \subset \left\{l_k^{\star}\right\} \cup \mathcal{A}_k\left[0\right]$ as the serving set of APs to UE $k$.
        \ELSE
        \STATE \label{UE_denied} UE $k$ is denied access to the network.
        \ENDIF
        \ENDFOR
        \STATE Update $\mathcal{D}_l\left[0\right], l = 1, \hdots, L$ based on the entries of $\mathcal{M}_k\left[0\right], k = 1, \hdots, K$.
\end{algorithmic}
\textbf{Output:} The initial AP cluster for each UE $\mathcal{M}_k\left[0\right], \forall k$ and UEs served by each AP $\mathcal{D}_l\left[0\right], \forall l$.
\end{algorithm}

\section{Soft Handover Procedures}\label{handover_sec}

In this section, we describe the proposed pilot assignment and cluster update algorithms that effectively handle the changes in the set of serving APs under mobility, as well as the assigned pilot to each UE. The AP-UE associations rely on the average channel gains between the APs and UEs. The constantly developing UE locations due to mobility result in a change of the channel gains between the APs and UEs in the network, and consequently the AP-UE associations.
As stated earlier, we let the parameter $T_i$ denote the shortest time period in which substantial changes in the average channel gains can occur due to UE mobility.
Hence, we present the handover procedures for updating the AP-UE associations at discrete time stamps that are equally spaced with the time interval $T_i$. At each discrete time instance, we first check on the AP cluster of each UE such that APs having average channel gains that are below the noise floor to a given UE are removed from the serving set of the UE and the reserved pilot sequence at those APs is released (i.e., lost connections). The set of UEs served by each AP is updated accordingly. In the rare occasion that a given UE loses connection to all APs in its serving set at the same time interval $n$, the UE is added to the set $\mathcal{F}\left[n\right]$ of UEs currently without connection.
If a UE loses connection to its master AP, the AP currently having the strongest average channel gain is appointed as the new master AP. Master AP handovers are primarily initiated by two events. The first is when the current master AP suffers from signal blockage due to a UE being masked by an obstruction and the second is when a UE is moving away from its master AP such that there is a gradual decrease in the average channel gain. The situation pointed out here represents the former case, which is important in mmWave scenarios.

We define two procedures for pilot assignment and cluster update; Algorithm~\ref{alg2} is specific to UEs currently without connection, i.e., $\mathcal{F}\left[n\right]$, while Algorithm~\ref{alg3} is for the remaining UEs having non-empty serving sets of APs. In either case, the previous assignment and clustering is taken as input, which could have been designed arbitrarily. In the simulations, we use the output of the initial access algorithm in Algorithm~\ref{alg1} as input:
$\mathcal{M}_k\left[0\right]$, $k = 1, \hdots, K$, and $\mathcal{D}_l\left[0\right]$, $l = 1, \hdots, L$.
Undesired ping-pong handovers due to small-scale channel variations can be combated by a straightforward adoption of the time-to-trigger (TTT) concept that is commonly used in the handover procedures of current cellular networks. Accordingly, TTT is not the focus of this work. We instead consider handover procedures that are triggered by large-scale UE mobility, since pilot assignment and cluster update represent the primary difference between cell-free and conventional cellular networks, from a handover algorithm perspective.

\subsection{Reconnecting UEs}

The pilot assignment and cluster update algorithm for reconnecting UEs is devised in Algorithm~\ref{alg2}. Each UE keeps on attempting to connect to an AP in descending order of perceived average channel gain between the UE and all APs in its vicinity (Steps~\ref{repeat} through \ref{until}). The first AP to accept the connection request (based on having a free service slot, i.e., $|\mathcal{D}_{O_{k}^{\left(n\right)}\left(i\right)}| < \tau_p$) is appointed as the \textit{Master} AP for that UE. Since the UE is attempting to reconnect to the network without having a current serving AP cluster, the procedure for the set of reconnecting UEs $\mathcal{F}\left[n\right]$ continues, in Step \ref{IAsteps}, as the initial access algorithm.

\begin{algorithm}[h!]
	\caption{Reconnecting UEs}
	\label{alg2}
	\noindent\textbf{Input:} For AP $l, l = 1, \hdots, L$, the remaining set of served UEs $\mathcal{D}_l\left[n\right]$.
	\begin{algorithmic}[1]
		\FOR{$k \in \mathcal{F}\left[n\right]$} \label{2part2}
            \STATE Sort the APs in descending order based on $\beta_{kl}\left[n\right], l = 1, \hdots, L$. \label{APsort}
		\STATE Set $\mathcal{A}_k\left[n\right] \gets \left\{O_{k}^{\left(n\right)}\left(1\right), \hdots, O_{k}^{\left(n\right)}\left(M_{k,\mathrm{min}}^{\left(n\right)}\right)\right\}$.
		\STATE Set $i \gets 0$.
            \REPEAT \label{repeat}
            \STATE Set $i \gets i+1$.
		\STATE UE $k$ sends a connection request to AP $O_{k}^{\left(n\right)}\left(i\right)$.
		\STATE Set $\mathcal{A}_k\left[n\right] \gets \mathcal{A}_k\left[n\right]\backslash\{O_{k}^{\left(n\right)}\left(i\right)\}$.
            \UNTIL $O_{k}^{\left(n\right)}\left(i\right)$ accepts service to UE $k$ \textbf{or} $i = M_{k,\mathrm{min}}^{\left(n\right)}$.\label{until}
		\IF{$O_{k}^{\left(n\right)}\left(i\right)$ accepts service to UE $k$}
		\STATE \label{IAsteps} Perform Steps~\ref{master_appoint} through \ref{UE_denied} in Algorithm~\ref{alg1} for time index $n$.
		\ENDIF
		\ENDFOR \label{2part2end}
            \STATE Update $\mathcal{D}_l\left[n\right],\forall l$ based on the entries of $\mathcal{M}_k\left[n\right], k \in \mathcal{F}\left[n\right]$.
		\end{algorithmic}
\textbf{Output:} The AP cluster $\mathcal{M}_k\left[n\right], k \in \mathcal{F}\left[n\right]$ and UEs served by each AP $\mathcal{D}_l\left[n\right],\forall l$.
\end{algorithm}

\subsection{Pilot Assignment and Cluster Update}

For the rest of the UEs having non-empty serving AP clusters, the proposed pilot assignment and cluster update at time interval $n$ is devised in Algorithm~\ref{alg3}.
Since the set of serving APs is gradually updated as a UE moves along a trajectory, Algorithm~\ref{alg3} acts as a soft handover procedure that revises the serving AP set to remain suitable for the UE's current location. A main goal of the algorithm is to reduce the number of master AP handovers, necessary pilot assignment decisions and pilot contamination, while maintaining good SE performance for all the UEs in the network with distributed operation and reasonable computational complexity. To achieve this, the algorithm is designed such that a minimal number of pilot changes (which affects the update of the serving set of APs) is required for each UE. Moreover, similar to the initial access algorithm, pilot sharing is only allowed for UEs having disjoint AP clusters, limiting the pilot contamination in the network.

Similar to the case of reconnecting UEs, in Step~\ref{2part3} each UE first identifies the AP with the strongest average channel gain, conditioned on that AP having a free slot to serve the UE; that is $O_{k}^{\left(n\right)}\left(i\right)$. If such an AP exists, we define $\beta_{\mathrm{max}}$ and $\beta_{\mathrm{curr}}$ as the average channel gains of $O_{k}^{\left(n\right)}\left(i\right)$ and the current master AP of the UE, respectively.
Next, if $\beta_{\mathrm{max}}$ is larger than $\beta_{\mathrm{curr}}$ with a proper handover margin $M_{\mathrm{HO}}$, then $O_{k}^{\left(n\right)}\left(i\right)$ is appointed as the new master AP of UE $k$, denoted as $l_k^{\star}$ (Step~\ref{new_master}). Moreover, if the newly appointed master AP $l_k^{\star}$ neither belongs to the current serving set of UE $k$ nor has the pilot already assigned to the UE free, a new pilot assignment decision and cluster formation needs to be performed as described in Steps~\ref{pilot_int} through \ref{M_ksetting} in Algorithm~\ref{alg1}. Otherwise, UE $k$ keeps the previously assigned pilot sequence and $l_k^{\star}$ sends service invitations to the reported set of new APs that are not part of the AP cluster of UE $k$; that is $\mathcal{A}_k\left[n\right] \backslash \mathcal{M}_k\left[n\right]$.

\begin{algorithm}[t]
        \caption{Pilot Assignment and Cluster Update}
	\label{alg3}
	\noindent\textbf{Input:} For UE $k, k \in \{1, \hdots, K\} \backslash \mathcal{F}\left[n\right]$, the remaining AP cluster $\mathcal{M}_k\left[n\right]$. For AP $l, l = 1, \hdots, L$, the set of served UEs $\mathcal{D}_l\left[n\right]$.
        \begin{algorithmic}[1]
        \FOR{$k \in \{1, \hdots, K\} \backslash \mathcal{F}\left[n\right]$}
		\STATE Perform Steps \ref{APsort} through                \ref{until} in Algorithm~\ref{alg2}. \label{2part3}
		\IF{$O_{k}^{\left(n\right)}\left(i\right)$ accepts service to UE $k$}\label{condition}
		\STATE Define $\beta_{\mathrm{max}}$ as the average channel gain of $O_{k}^{\left(n\right)}\left(i\right)$.
		\STATE Define $\beta_{\mathrm{curr}}$ as the average channel gain of the current master AP.
		\IF{$\beta_{\mathrm{max}} > \beta_{\mathrm{curr}} + M_{\mathrm{HO}}$}
		\STATE \label{new_master} Set $l_k^{\star} \gets O_{k}^{\left(n\right)}\left(i\right)$.
		\STATE UE $k$ reports remaining set of serving APs $\mathcal{M}_k\left[n\right]$ to $l_k^{\star}$.
		\IF{$l_k^{\star} \notin \mathcal{M}_k\left[n\right] \And t_k \notin \mathcal{T}_{l_k^{\star}}$}
		\STATE Perform Steps~\ref{pilot_int} through \ref{M_ksetting} in Algorithm~\ref{alg1} for time index $n$.
		\ELSE
		\STATE \label{2UE_reports} UE $k$ reports the set $\mathcal{A}_k\left[n\right]$ to $l_k^{\star}$.
		\STATE $l_k^{\star}$ sends a service invitation on pilot $t_k$ to $\mathcal{A}_k\left[n\right] \backslash \mathcal{M}_k\left[n\right]$.
		\STATE An AP in $\mathcal{A}_k\left[n\right] \backslash \mathcal{M}_k\left[n\right]$ accepts invitation $\iff$ Pilot $t_k$ is free on that AP.
		\STATE Add APs that accepted the invitation to $\mathcal{L}_k\left[n\right]$.
		\STATE \label{2strongest} Update $\mathcal{M}_k\left[n\right] \gets$ strongest $\min\left(M_{\mathrm{max}},|\mathcal{M}_k\left[n\right] \cup \mathcal{L}_k\left[n\right]|\right) \in \left(\mathcal{M}_k\left[n\right] \cup \mathcal{L}_k\left[n\right]\right)$.
		\ENDIF
		\ELSE
		\STATE \label{same_steps} Perform Steps~\ref{2UE_reports} through \ref{2strongest}.
		\ENDIF
		\ENDIF
		\ENDFOR
        \STATE Update $\mathcal{D}_l\left[n\right],\forall l$ based on the entries of $\mathcal{M}_k\left[n\right], k \in \{1, \hdots, K\} \backslash \mathcal{F}\left[n\right]$.
\end{algorithmic}
\textbf{Output:} The AP cluster for each UE $\mathcal{M}_k\left[n\right], k \in \{1, \hdots, K\} \backslash \mathcal{F}\left[n\right]$ and UEs served by each AP $\mathcal{D}_l\left[n\right], \forall l$.
\end{algorithm}

As before, an AP accepts the invitation if and only if it has the pilot assigned to the UE free. The AP cluster for UE $k$ is then updated according to Step~\ref{2strongest}, where the strongest APs in the current serving set $\mathcal{M}_k\left[n\right]$ and the set of APs that accepted the invitation $\mathcal{L}_k\left[n\right]$ are chosen to serve UE $k$. If there is no master AP change required for the UE, then it implies that no pilot change is required and accordingly the same steps (Steps~\ref{2UE_reports} through \ref{2strongest}) are performed in this case (line \ref{same_steps}). 
If the condition in Step~\ref{condition} is not met, the UE keeps the remaining set of APs from the previous interval and no change in the serving set is required since there are no more APs that can be added to the UE's current serving set. Finally, the set of UEs served by each AP at time instant $n$, $\mathcal{D}_l\left[n\right], l = 1, \hdots, L$, are updated based on the entries of $\mathcal{M}_k\left[n\right], k = 1, \hdots, K$.

Note that the proposed Algorithm~\ref{alg3} for updating the pilot assignment and clusters limits the necessary pilot changes to the case when there is a master AP handover such that the newly appointed master AP does not have the pilot previously assigned to the UE free. The idea is to enable the UE to keep its current serving set of APs and reduce the required overhead in the network that is needed for communication between the master AP and other APs in the current serving set, as well as reducing new service invitations that are required upon pilot change. The proposed algorithm attempts at limiting the number of essential changes in terms of pilot assignment and serving set of APs to each UE. Moreover, the greedy nature of the proposed algorithm alongside the APs' decision to accept service invitations if and only if the chosen pilot sequence for a given UE is free at that AP ensures that upon deciding on the pilot assignment and serving set of APs to a given UE, there are no changes to be performed for other UEs.

\section{Congestion Mitigation: Serving Set Based Pilot Assignment}\label{SSB}

As stated earlier, pilot sharing between UEs with overlapping AP clusters is not allowed in the proposed algorithm to limit pilot contamination within the network. As a result, the maximum number of UEs that can be simultaneously served by an AP is limited to $\tau_p$.
In highly loaded scenarios, the available service slots (free pilot sequences) at every AP are limited compared to the number of UEs in the network. Logically, as the number of UEs connected to the network increases, the available service slots at the APs decrease. The basic pilot assignment scheme, which may be the same as that employed by a cellular network, does not exploit the benefit of having multiple APs jointly serving the UEs in the pilot assignment decision. In other words, the decision is solely based on the interference sensed on a given pilot sequence at the master AP, with no regard to other APs that may possibly be available for service only on a subset of the pilots. To address this issue, we propose a modified pilot assignment strategy, namely the \textit{Serving Set Based} (SSB) pilot assignment.
The key idea is that UE $k$ is assigned a pilot according to
\begin{equation}
    t_k = \argmax_{t \in \mathcal{T}_{l_k^{\star}}}\left( \sum_{l \in \mathcal{A}_k^{\prime}} \beta_{kl}\left[n\right]b_{tl}\left[n\right] - \sum_{i \in \mathcal{P}_t} \beta_{il_k^{\star}}\left[n\right]\right),
    \label{modifed_best_pilot}
\end{equation}
where $\mathcal{A}_k^{\prime}$ represents the $\min(M_{\mathrm{max}},M_{k,\mathrm{min}}^{(n)})$ APs having the strongest average channel gains to UE $k$ at a given time instant $n$ and $b_{tl}\left[n\right]$ is a binary indicator given by
\begin{equation}
    b_{tl}\left[n\right] = \begin{cases}
          1 & \quad\textrm{if pilot $t$ is free on AP $l$,}\\
          0 & \quad\textrm{otherwise.}
    \end{cases}
\end{equation}

The average channel gain information of APs in the vicinity of the UE is assumed to be readily available at the master AP. The assumption is deemed practical as UE measurement reports in current cellular networks already include information about received signal strengths from nearby APs in the so called \textit{Candidate Set}, which has been denoted earlier by $\mathcal{A}_k\left[n\right]$, having average channel gains to a given UE satisfying a predetermined threshold.

\begin{figure}
\centering
\setlength{\abovecaptionskip}{0.3cm}
\hspace{-1em}\includegraphics[scale=0.48]{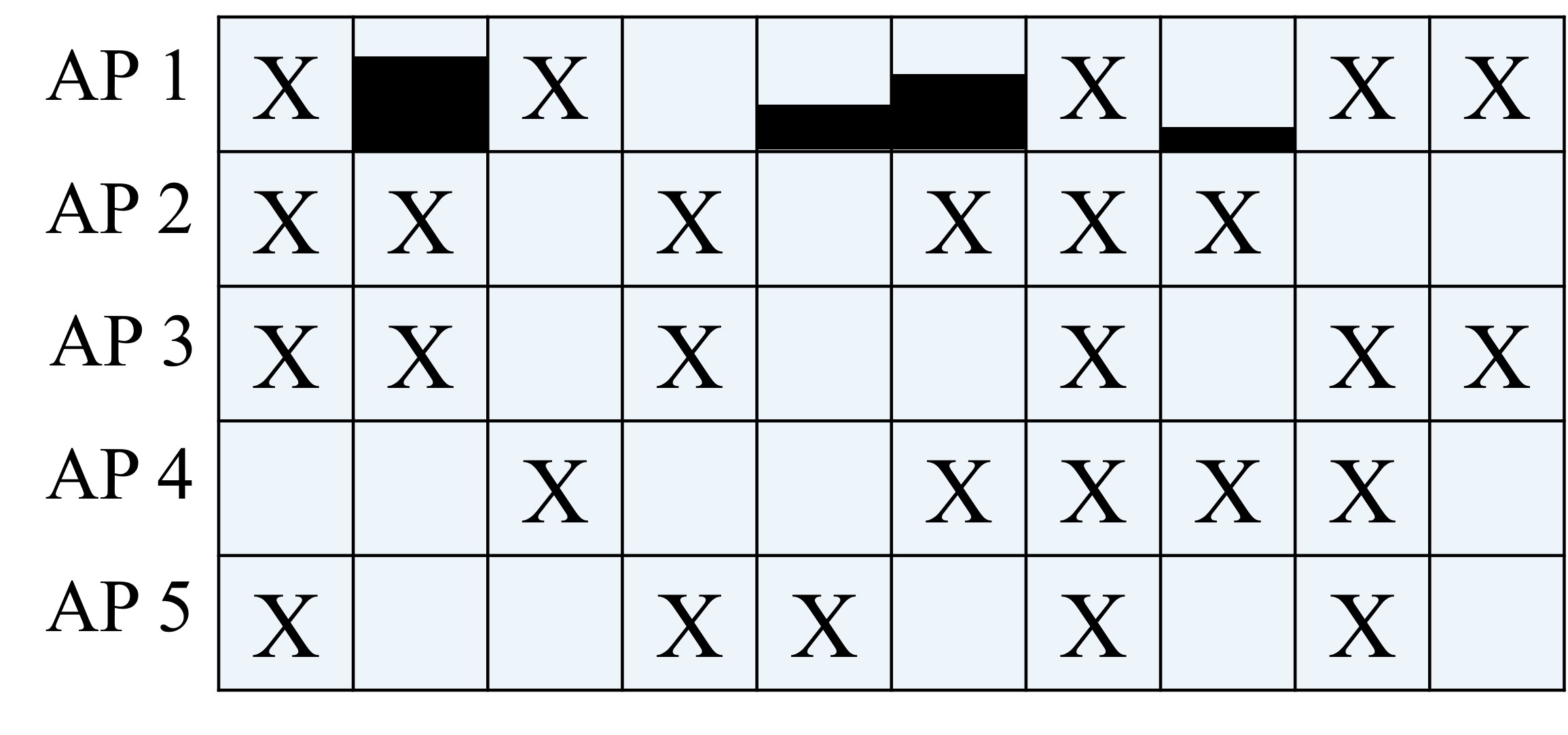}
\caption{An example of pilot assignment in a loaded cell-free network. Each row represents the $\tau_p$ pilot sequences at an AP. Locations marked with $\left(\textbf{X}\right)$ represent occupied pilots. Shaded blocks are pilots where interference is sensed at the master AP. SSB strategy balances between the interference at the master AP and the availability of other APs near a given UE.}
\label{model2}
\vspace{-1em}
\end{figure}

To illustrate the plausible advantage of the SSB pilot assignment strategy, especially in highly loaded scenarios, Fig.~\ref{model2} presents an example with $5$ APs that may possibly provide service to a given UE. The APs are sorted in descending order of the average channel gains to that UE and the locations marked with $\left(\textbf{X}\right)$ represent occupied pilot sequences at the APs. The basic pilot assignment scheme focusing solely on the master AP will always result in choosing pilot $4$ where no interference is sensed at the master AP, and accordingly depriving the UE of most of the APs in its vicinity. On the other hand, the SSB strategy uses a low-complex metric to balance the interference at the master AP and the availability of the chosen pilot sequence at other possible serving APs. In cases where the UE has comparable average channel gains to several APs (e.g., when it is in the middle between those APs), the possible serving APs that are available on a given pilot dominate the SSB metric. Alternatively, when the UE is relatively close to the master AP, minimizing the interference at the master AP may be more important. In Section \ref{numerical}, we demonstrate the performance improvement of the SSB pilot assignment over the basic scheme in highly loaded scenarios.

\section{Numerical Evaluation}\label{numerical}

In this section, we evaluate the performance of the proposed distributed pilot assignment and cluster update algorithm under UE mobility. We compare the performance of the UC cell-free massive MIMO network with that of a cellular UDN with the same AP locations, however, with each UE being served by a single AP. The communication performance is measured in terms of DL achievable SEs with MR and RZF precoding, based on the framework described in Section~\ref{system}.

\subsection{Mobility Model}\label{mobility}

The modeling of a UE's movement is a crucial building block for performance evaluation of the algorithms described in Section~\ref{handover_sec}. The choice of the mobility model has a direct effect on the obtained results. We consider a \textit{microscopic} mobility model where the UE movement is described by its space and speed coordinates at given time instants, which is deemed the most suitable modeling approach for the cell-free network topology consisting of many low-power APs with no defined rigid boundaries \cite{bettstetter2001smooth}. In the following, we present the mobility model developed to test the proposed pilot assignment and cluster update algorithm.

Inspired by the widely adopted \textit{Smooth Random Waypoint Mobility} model \cite{bettstetter2001smooth}, we develop a modified algorithm with maneuvering capabilities suitable for simulation of realistically moving UEs within a site map that includes buildings and other obstructions. The modifications lie in the choice of the UEs' initial and target locations as well as their directions of movement, which are restricted by the site geometry and obstacles along the UEs' trajectories. First, we denote the UEs' current and target locations at a given time instant $n$ by $(x_k^{\left(n\right)}, y_k^{\left(n\right)})$ and $(x_{k, \mathrm{t}}^{\left(n\right)}, y_{k, \mathrm{t}}^{\left(n\right)})$, $k = 1, \hdots, K$, respectively. We assume that all UEs move along the same horizontal plane at a fixed height of $h_{\mathrm{UE}}$ above the ground. Moreover, we denote the directions of movement of the UEs by $\theta_k^{\left(n\right)} \in \left[0, 2\pi\right]$, $k = 1, \hdots, K$, the number of steps that a UE takes while changing its direction of movement as $h$ and the remaining steps for UE $k$ upon direction change by $h_k$. Note that the concept of steps pointed out here represent the number of time intervals that a UE takes to change its direction of movement to avoid sharp turns in the UEs' position traces, as we consider a more realistic \textit{smooth} mobility model. The site map is divided into pixels of size $1\,\mathrm{m} \times 1\,\mathrm{m}$ and a flag map marking the locations of obstructions is defined as
\begin{equation}
    \mathbf{S}: \quad\mathbf{S}_{ij} = \begin{cases}
          0 & \quad\textrm{if obstruction,}\\
          1 & \quad\textrm{if free location.}
    \end{cases}
\end{equation}
where $\mathbf{S}_{ij}$ denotes location $\left(i, j\right)$, $i \in \left[1, x_{\mathrm{max}}\right]$, $j \in \left[1, y_{\mathrm{max}}\right]$ and $x_{\mathrm{max}}$, $y_{\mathrm{max}}$ represent the boundaries of the map. Finally, we define the random variables $r \sim \mathcal{U}\left[0, 1\right]$ and $d_{\mathrm{seg}} \sim \mathcal{U}\left[d_{\mathrm{min}}, d_{\mathrm{max}}\right]$, the maximum scanning angle $\theta$, and $d_s$ which represents the distance moved, in meters, by a UE within a single time interval. A summary of the mobility model is described in Algorithm~\ref{alg4}. Note that $\mathbf{nint}\left(\cdot\right)$ corresponds to rounding to the nearest integer value.

\begin{algorithm}[t!]
	\caption{Maneuvering Smooth Random Waypoint Mobility}
	\label{alg4}
	\noindent\textbf{Input:} Initial UE locations $\left(x_k^{\left(0\right)}, y_k^{\left(0\right)}\right)$, target locations $\left(x_{k,t}^{\left(0\right)}, y_{k,t}^{\left(0\right)}\right)$ and corresponding directions of movement $\theta_k^{\left(0\right)}$. Initialize $h_k = 0, k = 1, \hdots, K$.
	\begin{algorithmic}[1]
	    \FOR{$n = 1, 2, 3, \hdots$}
		\FOR{$k = 1, \hdots, K$}
		\IF{$h_k \neq 0$}
		\STATE $\theta_{\mathrm{temp}} \gets \left(\theta_k^{\left(n\right)} - \theta_k^{\left(n-1\right)}\right) / h_k$.
		\STATE $\theta_k^{\left(n\right)} \gets \theta_k^{\left(n-1\right)} + \theta_{\mathrm{temp}}$.
		\STATE $h_k \gets h_k - 1$.
		\ENDIF
		\STATE $\left(x_k^{\left(n\right)}, y_k^{\left(n\right)}\right) \gets \mathbf{nint}\left(x_k^{\left(n-1\right)} + d_s\cos\left(\theta_k^{\left(n\right)}\right),\right.$
  
        $\left. y_k^{\left(n-1\right)} + d_s\sin\left(\theta_k^{\left(n\right)}\right)\right)$.
		\STATE $i \gets 1$.
		\WHILE{$x_k^{\left(n\right)} \notin \left[1, x_{\mathrm{max}}\right] \hspace{0.5em} |\hspace{0.5em}| \hspace{0.5em} y_k^{\left(n\right)} \notin \left[1, y_{\mathrm{max}}\right] \hspace{0.5em} |\hspace{0.5em}| \hspace{0.5em} \mathbf{S}_{x_k^{\left(n\right)}y_k^{\left(n\right)}} = 0$}
		\STATE $h_k \gets 0$.
		\STATE $\theta_k^{\left(n\right)} \gets \angle\left(x_{k, t}^{\left(n-1\right)} - x_k^{\left(n-1\right)}, y_{k, t}^{\left(n-1\right)} - y_k^{\left(n-1\right)}\right)$.
		\IF{$\mathbf{mod}\left(i, 2\right) \neq 0$}
		\STATE $\theta_{\mathrm{temp}} \gets \theta_k^{\left(n\right)} + \left(\frac{i + 1}{2}\right)r\theta$.
		\ELSE
		\STATE $\theta_{\mathrm{temp}} \gets \theta_k^{\left(n\right)} - \left(\frac{i}{2}\right)r\theta$.
		\ENDIF
		\STATE $\left(x_k^{\left(n\right)}, y_k^{\left(n\right)}\right) \gets \mathbf{nint}\left(x_k^{\left(n-1\right)} + d_s\cos\left(\theta_{\mathrm{temp}}\right),\right.$
  
        $\left.y_k^{\left(n-1\right)} + d_s\sin\left(\theta_{\mathrm{temp}}\right)\right)$.
		\STATE $i \gets i + 1$.
		\ENDWHILE
		\vspace{2pt}\IF{$\norm{\left(x_{k, t}^{\left(n\right)} - x_k^{\left(n\right)}, y_{k, t}^{\left(n\right)} - y_k^{\left(n\right)}\right)} < d_s$}\vspace{2pt}
		\STATE $h_k \gets h$.
		\STATE $\left(x_{k,t}^{\left(n\right)}, y_{k,t}^{\left(n\right)}\right) \gets \left(0,0\right)$.
		\vspace{1pt}\WHILE{$x_{k,t}^{\left(n\right)} \notin \left[1, x_{\mathrm{max}}\right] \hspace{0.5em} |\hspace{0.5em}| \hspace{0.5em} y_{k,t}^{\left(n\right)} \notin \left[1, y_{\mathrm{max}}\right] \hspace{0.5em} |\hspace{0.5em}| \hspace{0.5em} \mathbf{S}_{x_{k,t}^{\left(n\right)}y_{k,t}^{\left(n\right)}} = 0$}
		\STATE $\theta_k^{\left(n\right)} \gets \theta_k^{\left(n\right)} + \left(r-0.5\right)\pi$.
		\vspace{1.5pt}\STATE $\left(x_{k,t}^{\left(n\right)}, y_{k,t}^{\left(n\right)}\right) \gets \mathbf{nint}\left(x_k^{\left(n\right)} + d_{\mathrm{seg}}\cos\left(\theta_k^{\left(n\right)}\right),\right.$
  
        $\left.y_k^{\left(n\right)} + d_{\mathrm{seg}}\sin\left(\theta_k^{\left(n\right)}\right)\right)$.
		\vspace{0.5pt}\ENDWHILE
		
		\ENDIF
            \vspace{1pt}\STATE $\left(x_{k,t}^{\left(n+1\right)}, y_{k,t}^{\left(n+1\right)}\right) \gets \left(x_{k,t}^{\left(n\right)}, y_{k,t}^{\left(n\right)}\right)$.
            \vspace{1pt}\STATE $\theta_k^{\left(n+1\right)} \gets \angle\left(x_{k, t}^{\left(n+1\right)} - x_k^{\left(n\right)}, y_{k, t}^{\left(n+1\right)} - y_k^{\left(n\right)}\right)$.
		\ENDFOR
        \ENDFOR
    \end{algorithmic}
    \textbf{Output:} Updated UE locations and target locations.
\end{algorithm}

\subsection{Simulation Setup}

We consider a cell-free network composed of $L = 64$ APs deployed in an area of $600$\,m $\times$ $600$\,m. Only the inner $500$\,m $\times$ $500$\,m are considered for generating the results, to reduce the boundary effects. All APs are equipped with half-wavelength-spaced ULAs. Moreover, we assume that $K \in \left[20,60\right]$ UEs are randomly and uniformly dropped within the area of interest, with each UE being given a uniform random initial target location, $x_{k,t}^{\left(0\right)}$, that lies within a distance of $d_{\mathrm{seg}} \sim \mathcal{U}\left[50, 100\right]$\,m from the initial UE location. The network operates at the typical mmWave frequency of $f_c = 28$\,GHz. 
The cell-free network simulation parameters are summarized in Table~\ref{params}. The scaling parameter $v$ in the power allocation policy in \eqref{input} is chosen to be $0.5$, which represents a good balance between total SE maximization and UE fairness. The UE mobility parameters utilized to generate all the results hereafter are summarized in Table~\ref{UEparams}. The results are averaged over $25$ random initial UE locations such that for each random drop, UEs are allowed to move over a duration of $400$ time intervals, i.e., $400\,T_i$ seconds, according to the mobility model described in Section~\ref{mobility}. We use the same platform for performing all simulations, a 4 core Intel(R) Core i5-10310U CPU with 1.7 GHz base frequency and 4.4 GHz max turbo frequency.

\subsection{Site Map and UE Mobility}

Most of the cell-free massive MIMO literature considers stochastic channel models for urban scenarios with static UEs; each UE drop experiences different AP locations and objects in the propagation environment (only modelled through shadow fading).
Such models cannot be used for meaningful evaluation of handover procedures due to large-scale UE mobility.
To capture the key characteristics of the realistic propagation environments, we use the ray-tracing software in \cite{simic2017demo} to obtain a spatially consistent channel model.
The APs are assumed to be deployed at a height of $6$\,m, whereas UEs move along the same horizontal plane at a height of $h_\mathrm{UE} = 1.5$\,m.
The APs deployment height corresponds to ULAs placed on utility poles and building facades. $h_\mathrm{UE}$ and $v_\mathrm{UE}$ are selected to represent handheld devices by pedestrians.

\begin{table}
      \centering
        \caption{Network simulation parameters.}
        \begin{tabular}{ll}
        \hline
        \textbf{Parameter} & \textbf{Value} \\
        \hline
        Area of interest & $500$\,m $\times$ $500$\,m \\
        Carrier frequency & $f_c = 28$\,GHz \\
        Bandwidth & $500$\,MHz \\
        Number of APs & $L = 64$ \\
        Number of UEs & $K \in \left[20,60\right]$ \\
        Number of antennas per AP & $N = 8$ \\
        Per-AP maximum DL transmit power & $P_{\textrm{max}}^{\textrm{dl}} = 1$\,W \\
        UL transmit power & $p_i = 100$\,mW \\
        UL/DL noise power & $-80$\,dBm \\
        Coherence block length & $\tau_c = 200$ \\
        Pilot sequence length & $\tau_p = 10$ \\
        \hline
        \label{params}
        \end{tabular}
\end{table}
\begin{table}
      \centering
        \caption{UE mobility parameters.}
        \begin{tabular}{ll}
        \hline
        \textbf{Parameter} & \textbf{Value} \\
        \hline
        Direction change steps & $h = 3$ \\
        Segment length & $d_{\mathrm{seg}} \sim \mathcal{U}\left[50,100\right]$\,m \\
        UE speed & $v_{\mathrm{UE}} \in \left[1.5, 6\right]$\,m/s \\
        Distance moved in one $T_i$ & $d_s = 0.75$\,m \\
        Simulation sampling time & $T_i = d_s/v_{\mathrm{UE}}$ \\
        Maximum scanning angle & $\theta = 18.4\degree$ \\
        \hline
        \end{tabular}
        \label{UEparams}
    \vspace{-0.5em}
\end{table}

\begin{figure}[h!]
\centering
\setlength{\abovecaptionskip}{0.24cm}
\includegraphics[scale=0.472]{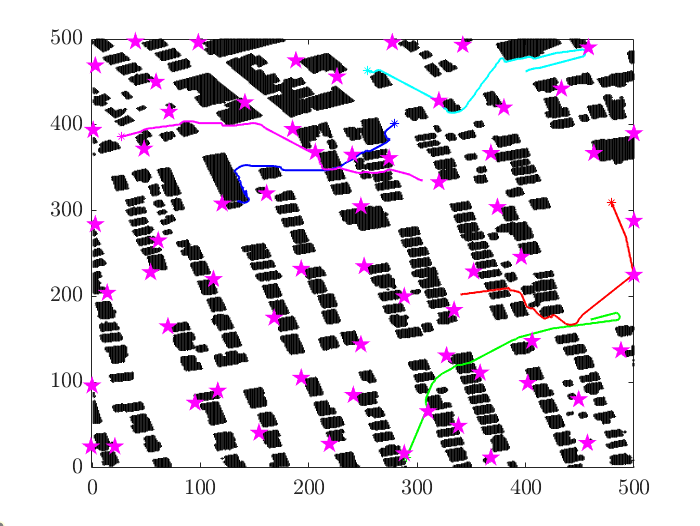} \vspace{-2mm}
\caption{Site map with $K = 5$ moving UEs over a $400\,T_i$ time interval.}
\label{UE_mobility}
\vspace{-1em}
\end{figure}

To demonstrate the feasibility of the developed random mobility model, Fig.~\ref{UE_mobility} presents the simulated site map with buildings/obstructions and $K = 5$ moving UEs over a duration of $400\,T_i$ seconds. The site map and buildings' sizes correspond to typical European cities. It can be seen that the mobility model results in reasonable UE position traces that simulate, to a great extent, countless possible situations that the cell-free network may be subject to in practice.

\subsection{Achievable Spectral Efficiency}

\begin{figure}
\setlength{\abovecaptionskip}{0.3cm plus 0pt minus 0pt}
\begin{subfigure}{.5\textwidth}
\centering
\includegraphics[scale=0.5]{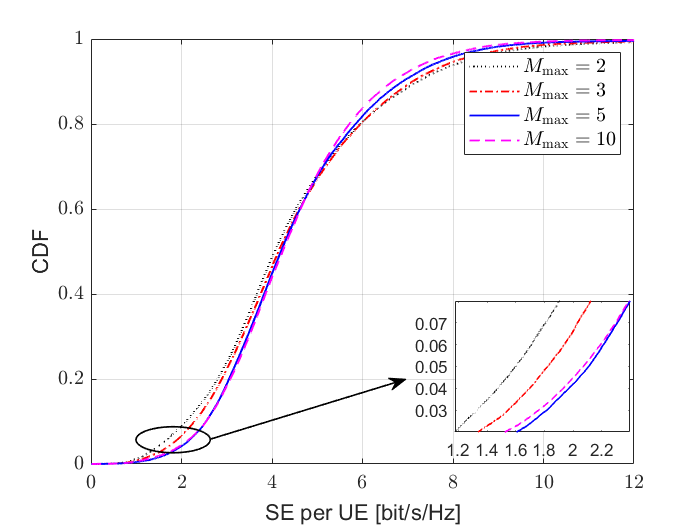}
\caption{$K = 20$.}
\label{MR_K=20}
\end{subfigure}
\begin{subfigure}{.5\textwidth}
\centering
\includegraphics[scale=0.5]{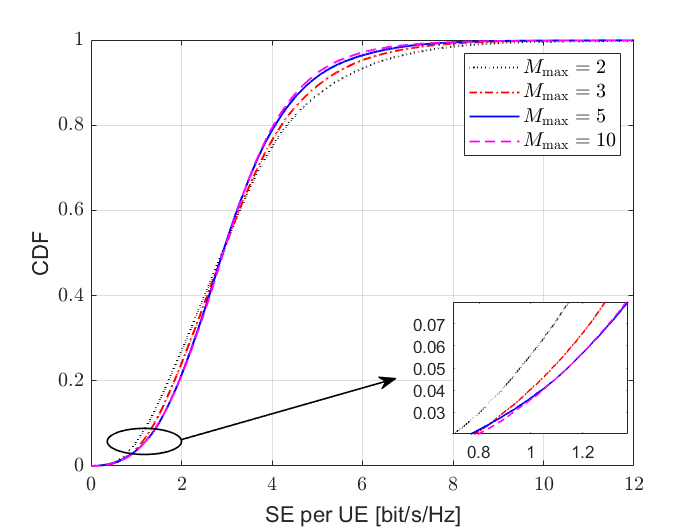}
\caption{$K = 50$.}
\label{MR_K=50}
\end{subfigure}
\caption{CDF of the DL SE per UE with MR precoding and different maximum cluster sizes.}
\label{MR}
\vspace{-1em}
\end{figure}

In this subsection, we evaluate the cell-free massive MIMO network performance in terms of the achievable SE per UE with MR and RZF precoding. First, we compare the performance of the system considering different maximum allowable AP cluster sizes serving any given UE. The ultimate motivation behind the cell-free network topology is to eliminate poor cell-edge performance that is inherent to conventional cellular setups and provide uniform and reliable service to all UEs. For that reason, we focus on highlighting the $95\%$-likely SE performance for a given UE which is considered a primal indicator of the performance of cell-free networks. All results in this section are generated using $v_{\mathrm{UE}} = 1.5$\,m/s. Note that SE variations resulting from ageing of the channel estimates due to UE mobility are not considered in our model, since the SE estimate is computed at the same time stamps as the clusters, pilots and channels are updated. Such variations are expected to be minor for the considered low-speed dense urban scenario where cell-free networks and mmWave band operation is most suitable. Fig.~\ref{MR} plots the cumulative distribution function (CDF) of the SE per UE with MR precoding and different cluster sizes for $K = 20$ and $K = 50$ moving UEs. The SE variations are representative for what UEs will experience when moving around in the area.
The $95\%$-likely SE shows a tangible performance gain, $32\%$, when increasing $M_{\mathrm{max}}$ from $2$ to $5$ APs, especially for the case with $K = 20$ UEs. For $K = 50$ UEs, the results encourage the usage of the same maximum cluster size, where increasing $M_{\mathrm{max}}$ beyond $5$ APs does not offer any performance improvement for the simulated scenarios. Moreover, it can be seen that the $95\%$-likely SE per UE reduces by roughly $40\%$ compared to the case of $K = 20$ UEs due to the added interference. However, when comparing the total DL SE, a substantial increase is noted with increasing $K$ in the range $\left[20, 50\right]$ UEs since the SE loss per UE is relatively small.

\begin{figure}
\setlength{\abovecaptionskip}{0.3cm plus 0pt minus 0pt}
\begin{subfigure}{.5\textwidth}
\centering
\includegraphics[scale=0.5]{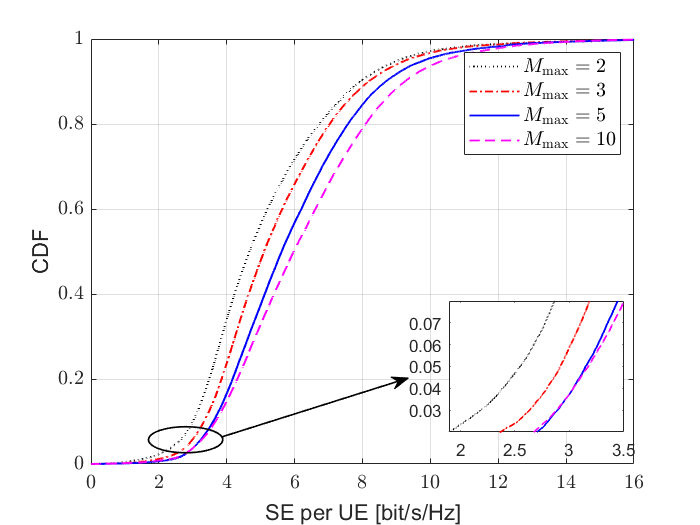}
\caption{$K = 20$.}
\label{RZF_K=20}
\end{subfigure}
\begin{subfigure}{.5\textwidth}
\centering
\includegraphics[scale=0.5]{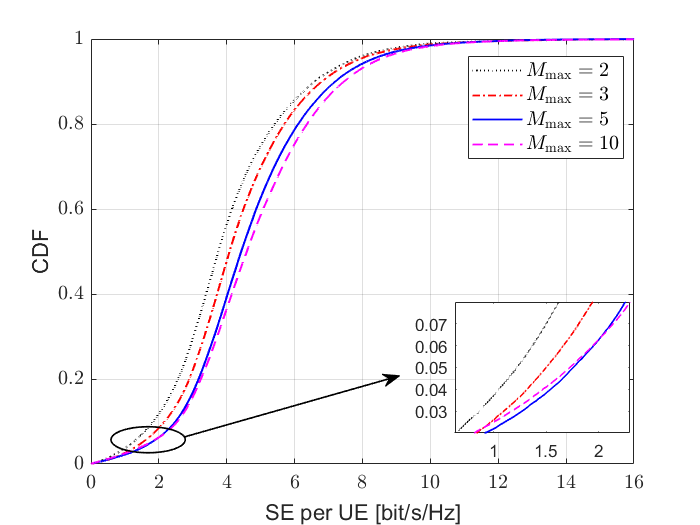}
\caption{$K = 50$.}
\label{RZF_K=50}
\end{subfigure}
\caption{CDF of the DL SE per UE with RZF precoding and different maximum cluster sizes.}
\label{RZF}
\vspace{-1em}
\end{figure}

Fig.~\ref{RZF} shows the CDF of the SE per UE with RZF precoding and different cluster sizes for $K = 20$ and $K = 50$ UEs. Similar behaviour to the case of MR precoding is observed suggesting the maximum cluster size that is needed in this setup is still  $M_{\mathrm{max}} = 5$. However, the performance improvement, in [bit/s/Hz], achieved by increasing $M_{\mathrm{max}}$ from $2$ to $5$ APs is higher than with MR. The reason behind this is that RZF precoding attempts at suppressing the inter-user interference and thus can manage a higher number of AP-UE links operating at the same time with better performance. When increasing $M_{\mathrm{max}}$ beyond $5$ APs with RZF precoding, it is clear that a performance gain exists for more fortunate UEs with better channel conditions, especially for $K = 20$ UEs. However, no additional improvements are seen in terms of $95\%$-likely SEs, which is the key indicator of a cell-free network performance, as previously stated. The results reveal that for all the simulated scenarios with different numbers of UEs and precoding schemes, a good choice of the maximum AP cluster size serving a UE is $M_{\mathrm{max}} = 5$. Beyond that, the network suffers from increased signalling overhead between the APs (to support larger clusters) at no notable performance gain.

\begin{figure}
\setlength{\abovecaptionskip}{0.3cm plus 0pt minus 0pt}
\begin{subfigure}{.5\textwidth}
\centering
\includegraphics[scale=0.5]{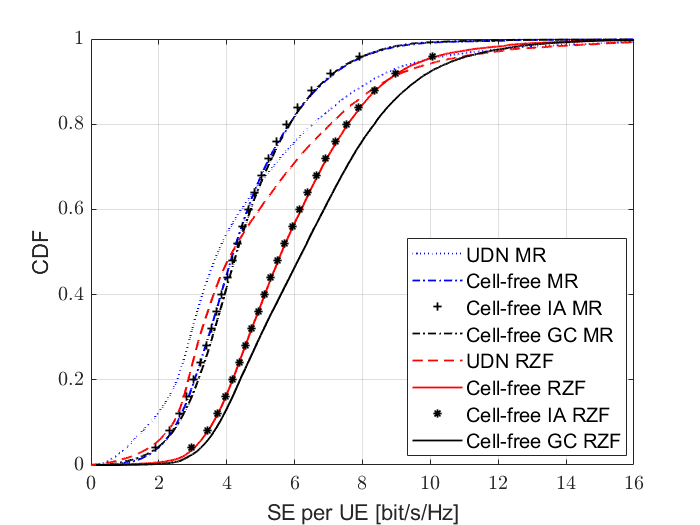}
\caption{$K = 20$.}
\label{CFvsUDN_K=20}
\end{subfigure}
\begin{subfigure}{.5\textwidth}
\centering
\includegraphics[scale=0.5]{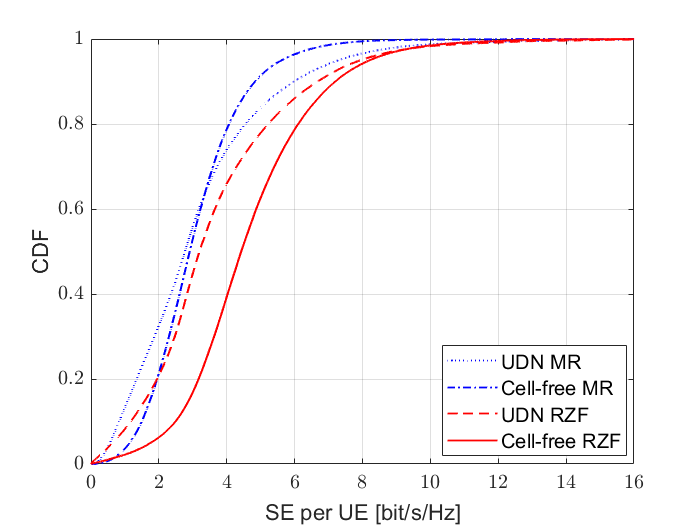}
\caption{$K = 50$.}
\label{CFvsUDN_K=50}
\end{subfigure}
\caption{CDF of the DL SE per UE for cell-free and UDN with MR and RZF precoding, $M_{\mathrm{max}} = 5$.}
\label{CFvsUDN}
\vspace{-1em}
\end{figure}

To evaluate the performance of our proposed pilot assignment and cluster update algorithm, Fig.~\ref{CFvsUDN} compares the CDF of the SE per UE of the proposed dynamic algorithm with that of a UDN, where each UE is served by a single AP, under MR and RZF precoding for $K = 20$ and $K = 50$ moving UEs. We further show the CDF of the SE per UE that is achieved by running the initial access algorithm, denoted by ``IA'', and that of the centralized optimal graph colouring (GC) benchmark in \cite{liu2020graph} at every time instance for $K = 20$ UEs. For the proposed algorithms, the maximum cluster size is set to $M_{\mathrm{max}} = 5$, based on the previous results. For the GC algorithm, there is no limit imposed on the cluster size, which is why it achieves a slightly higher SE with RZF for the more fortunate UEs. This can be achieved with a larger cluster size for our proposed algorithms (see Fig. \ref{RZF}), however, increasing the cluster size also imposes more signalling overhead. It can be seen that the proposed dynamic algorithm achieves the same performance as the IA algorithm. This is remarkable since the proposed algorithm makes local refinements to the pilot assignment and cluster formation triggered by UE mobility, while the IA benchmark updates everything for every UE at each time instance. Moreover, the gap between our proposed handover procedure and the centralized GC benchmark for RZF precoding is only $6\%$ in terms of the $95\%$-likely SE, whereas it is insignificant for MR precoding. This shows that our handover procedure identifies all the essential changes that need to be made to refine previous assignments, leading to nearly identical SE at a vastly reduced complexity. Note that the optimal GC benchmark cannot be utilized as a handover procedure for updating the serving clusters and pilots for each UE dynamically due to mobility as the algorithm optimizes the clusters and pilots for the UEs altogether, meaning that it requires redoing everything for all UEs whenever a handover event occurs for a single UE.
In addition, the proposed solution significantly outperforms the UDN architecture in terms of both $95\%$-likely SE and median SE, for both MR and RZF precoding. The performance difference is seen to be larger for the case of RZF due to its ability to mitigate the added interference that appears in a cell-free network with more serving APs per UE than in the UDN. When comparing Fig.~\ref{CFvsUDN_K=20} with Fig.~\ref{CFvsUDN_K=50}, we observe that the reduction in achievable SE per UE with increasing $K$ is smaller with the proposed solution compared to the UDN showing the superior capability of the proposed solution to handle a larger number of UEs with a given target SE per UE.

\subsection{SSB Pilot Assignment}

\begin{figure}
\centering
\setlength{\abovecaptionskip}{0.3cm}
\includegraphics[scale=0.5]{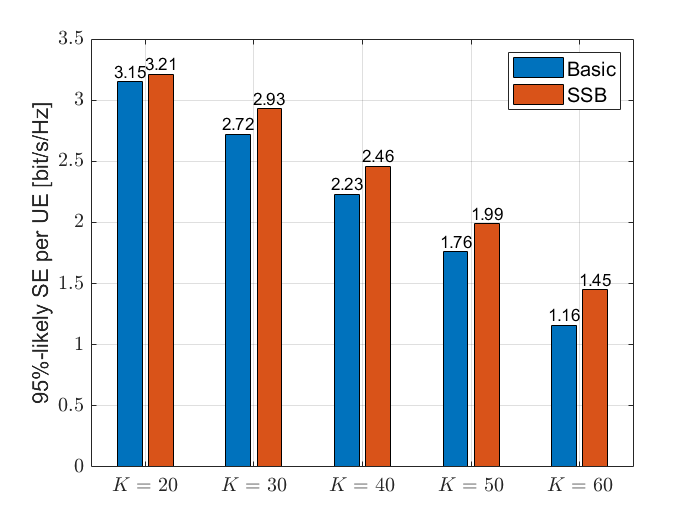}
\caption{$95\%$-likely SE per UE of the basic and SSB pilot assignments with RZF, $M_{\mathrm{max}} = 5$.}
\label{basicVSmodified}
\vspace{-1em}
\end{figure}

In this subsection, we compare the performance of the proposed SSB pilot assignment with that of the low-complexity original pilot assignment metric presented in Section \ref{pilot}. Fig.~\ref{basicVSmodified} plots the $95\%$-likely SE per UE with RZF precoding and the maximum cluster size $M_{\mathrm{max}} = 5$ to compare the basic and SSB pilot assignment schemes for $K \in \left[20, 60\right]$. We notice that the performance improvement of the SSB scheme over the basic scheme increases with $K$. For instance, the SE improvement is  $7.6\%$ with $K = 30$ UEs, whereas the improvement is  $13.3\%$ with $K = 50$ UEs. 
In the even more congested scenario of $K = 60$ UEs, the SE improvement reaches roughly $25\%$. Intuitively, a more challenging situation for pilot assignment and cluster formation arises when the number of simultaneously served UEs is much greater than the number of orthogonal pilot sequences, resulting in more pilot sequences being occupied at each AP. This necessitates a more involved pilot assignment procedure that takes into account the load distribution (used pilots) on nearby APs that are in the vicinity of a given UE. Note that with the modified SSB Pilot Assignment strategy, the maximum total DL SE is achieved at a higher number of UEs, i.e., the modified scheme improves the capability of the network to handle a larger number of UEs with better total SE performance.

For MR precoding, no tangible performance gain is observed with the modified scheme for the simulated scenarios, except for $K = 60$ UEs where the $95\%$-likely SE per UE recorded an increase from $0.94$ to $1.03$ [bit/s/Hz] with approximately $9.5\%$ improvement. The reason behind this is that pilot contamination has a more detrimental effect on RZF precoding compared to MR precoding.
Accordingly, the use of the lower complexity pilot assignment procedure is deemed sufficient with the low-complexity precoder, unless the number of UEs grows large.

\subsection{Handover Analysis}

\begin{figure}
\setlength{\abovecaptionskip}{0.3cm plus 0pt minus 0pt}
\begin{subfigure}{0.5\textwidth}
\centering
\includegraphics[scale=0.5]{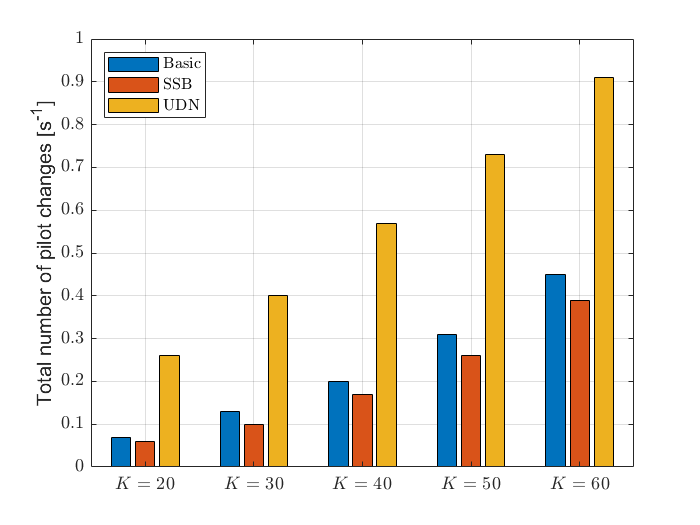}
\caption{$v_{\mathrm{UE}} = 1.5$\,m/s.}
\label{agaisntK}
\end{subfigure}
\begin{subfigure}{0.5\textwidth}
\centering
\includegraphics[scale=0.5]{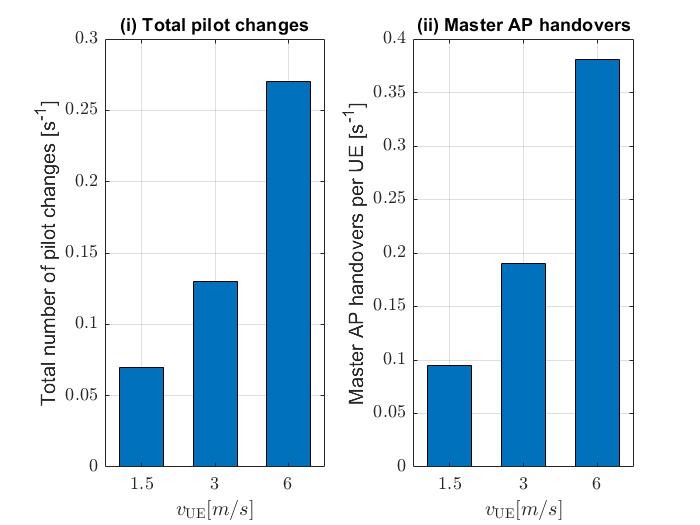}
\caption{$K = 20$ UEs.}
\label{agaisntV}
\end{subfigure}
\caption{Total pilot changes and master AP handovers for different $K$ and $v_{\mathrm{UE}}$, $M_{\mathrm{max}} = 5$.}
\label{pilot_change_bar_gragh}
\vspace{-1em}
\end{figure}

This subsection highlights the performance of the proposed algorithms in terms of average number of master AP handovers, pilot assignment decisions and average cluster size serving a given UE.
First, Fig.~\ref{agaisntK} plots the average pilot sequence change rate for all UEs in the network in comparison to the UDN topology. Surprisingly, the pilot change rate for the cell-free topology is lower than that of the UDN. The reason behind this is that the cell-free network allows cooperation between the APs, forming a serving set of APs to each UE. The APs in the vicinity of the UE are thus more likely to have the assigned pilot by the previous master AP already reserved for that UE. In addition, our proposed solution limits the necessary pilot changes to the case when there exists a master AP handover, such that the new master AP neither has the assigned pilot free nor belongs to the current serving set of the UE (meaning that it has the pilot already reserved for that UE).
Moreover, as a complementary benefit of the proposed SSB pilot assignment, it can be seen that there exists a noticeable reduction (about 15\% for $K = 50$ UEs) in the recorded total number of pilot changes among the UEs, which is due to the fact that the SSB strategy takes into consideration the occupied pilot sequences at the APs that are in the vicinity of a given UE. This results in a higher likelihood that APs with stronger average channel gains, having a greater chance of serving as the master AP in a future time interval, are able to join the serving set of a given UE.

Since master AP handovers are triggered mainly by large-scale UE mobility, Fig. \ref{agaisntV} shows the total pilot change rate and master AP handover rate against $v_{\mathrm{UE}}$. Logically, the master AP and pilot changes increase with increasing $v_{\mathrm{UE}}$. It can be seen that doubling the UE speeds approximately doubles the master AP handovers, whereas it provides a slightly higher than double the pilot changes. Moreover, we observed that the average number of master AP handovers per UE remains fairly constant when varying the number of UEs $K$ and maximum cluster size $M_{\mathrm{max}}$, and is roughly the same as the number of handovers in the UDN system.

Next, Table~\ref{avg_cluster} records the average cluster size for different numbers of UEs $K$ and different maximum allowable cluster sizes $M_{\mathrm{max}}$. It is observed that the average serving set size is not notably affected by increasing the number of UEs. This means that for most of the locations within the area of interest, there exists more than $5$ APs that may possibly provide service to a UE at that location, i.e., $M_{k,\mathrm{min}}^{\left(n\right)} \geq 5, \forall n, k$. Only a slight decrease for the case of $M_{\mathrm{max}}= 5$ is noted, which is due to more total number of service slots (pilot sequences) being occupied by the UEs in this case; resulting in a higher likelihood that the pilot assigned to a given UE is occupied at some of the possible serving APs for that UE.

\begin{table}
\begin{center}
\caption{Average number of APs serving a UE.}
\begin{tabular}{ |c|c|c|c|c|c|  }
\hline
$M_{\mathrm{max}}$ & $K = 20$ & $K = 30$ & $K = 40$ & $K = 50$ & $K = 60$ \\
\hline
$5$ & $4.41$ & $4.38$ & $4.28$ & $4.26$ & $4.16$\\
$3$ & $2.90$ & $2.88$ & $2.88$ & $2.87$ & $2.87$\\
\hline
\end{tabular}
\label{avg_cluster}
\end{center}
\vspace{-1em}
\end{table}

\subsection{Complexity Analysis And Signalling Overhead}

The complexity of the proposed algorithms lie mainly in the sorting operations. In addition, for each UE $k$, $k = 1, \hdots, K,$ the pilot assignment decision requires choosing a pilot from a set with cardinality of, at most, $\tau_p$. The other intermediate steps in each of Algorithms 1-3 represent a relatively negligible component to the complexity, which does not affect the order of $O(\cdot)$ operations. As such, the IA algorithm in Algorithm \ref{alg1} has a complexity of $O\left(L\left(K\mathrm{log}_2\tau_p\right) + K\left(L\mathrm{log}_2L + \tau_p\right)\right)$. For each UE $k \in \mathcal{F}\left[n\right]$, the complexity for Algorithm \ref{alg2}, that is specific to reconnecting UEs, is $O\left(L\mathrm{log}_2L + \tau_p\right)$. As for Algorithm \ref{alg3}, if a new pilot assignment decision is required, the complexity becomes the same as in Algorithm \ref{alg2}. Otherwise, the complexity for a given UE reduces to $O\left(L\mathrm{log}_2L\right)$. On the other hand, the centralized GC scheme has a complexity of $O\left(K\left(K + 2L\right) + KL\mathrm{log}_2L\right)$ \cite{liu2020graph}. However, the reported complexity in \cite{liu2020graph} does not account for the bisection search for updating the interference graph to find the optimal solution. To quantify the complete complexity, we supplement the analytic formulas with the recorded average run-time over $100$ samples for the case of $K = 50$ UEs in Table \ref{runtime}. It is clear that the proposed algorithms require orders-of-magnitude lower run-time compared to the centralized GC benchmark.

\begin{table}
\begin{center}
\caption{\centering Average computational time in seconds for $K = 50$ UEs.}
\begin{tabular}{ |c|c|c|  }
\hline
IA & Handover & GC \\
\hline
0.0037 & 0.0022 & 0.749 \\
\hline
\end{tabular}
\label{runtime}
\end{center}
\vspace{-1em}
\end{table}
As for the signalling, our proposed algorithms only require computations of local metrics at a given UE or AP with the local information measured at each of them. Signalling is then only needed for the purpose of sending and receiving the decisions on the pilot assignment and cluster update between the relevant entities (e.g., the master AP and other serving APs to a given UE). On the other hand, centralized schemes, as in \cite{liu2020graph}, rely on computing centralized metrics that include information from several locations within the network. This necessitates that several coefficients are required to be exchanged between the APs to first compute these metrics then communicate the resulting decisions, imposing increased signalling overhead. It is worth noting that the decisions need to be communicated at the handover rate, whereas communicating large-scale fading coefficients for computation of centralized metrics will be required at the rate of change of the coefficients which is expected to be much faster than the handover rate.

\section{Conclusions}\label{conc}

One of the previously overlooked features of cell-free networks is handover, where the AP cluster that serves a UE changes during mobility.
In this paper, we have proposed a distributed algorithm that gradually updates the AP clusters and pilot assignment under UE mobility. We focused on the DL with distributed MR or RZF precoding.
The main goal is to extend traditional handover concepts to the more complicated AP-UE associations of cell-free networks, with focus on mmWave scenarios with APs along streets. The proposed algorithm is designed to minimize the number of master AP handovers and revised pilot assignment decisions. Moreover, we propose an enhanced pilot assignment strategy that takes into account the load distribution on neighbouring APs, showing significant improvement in highly loaded scenarios. 
To numerically evaluate the algorithm, we developed a novel mobility model to simulate realistic UE mobility in a site map including buildings/obstructions. The numerical results showed that our proposed distributed algorithms outperform the cellular UDN architecture and provide a remarkable improvement in the $95\%$-likely SE, thereby delivering more stable SE under mobility.
The proposed algorithms identify the essential refinements since they deliver comparable SE to the case when the association is completely redone at every time instance, however, requiring orders-of-magnitude lower computational time compared to the state-of-the-art.
Moreover, the algorithms trigger a number of master AP handovers and changes in pilot assignment that is practically feasible. In conclusion, the combination of the cell-free massive MIMO architecture and mmWave bands (known for high bandwidth but poor channel reliability) lead to unprecedented data rates with high reliability and network coverage.

\section*{References}
\renewcommand{\refname}{ \vspace{-\baselineskip}\vspace{-1.1mm} }
\bibliographystyle{ieeetr}
\bibliography{papercites}

\end{document}